\documentclass{iucrjournals}
\nolinenumbers

\usepackage{multirow}

\title{Tomo-center: an AI-based rotation-axis center finder
for synchrotron micro- and nano-tomography}

\author[a]{Songyuan~Tang%
  \IUCrCemaillink{stang@anl.gov}%
  \IUCrOrcidlink{0000-0002-2324-1726}}%
\author[a]{Xiaoyang~Liu%
  \IUCrEmaillink{xiaoyang@anl.gov}%
  \IUCrOrcidlink{0000-0002-9326-2135}}%
\author[a]{Viktor~Nikitin%
  \IUCrEmaillink{vnikitin@anl.gov}%
  \IUCrOrcidlink{0000-0001-9999-169X}}%
\author[a]{Alberto~Mittone%
  \IUCrEmaillink{amittone@anl.gov}%
  \IUCrOrcidlink{0000-0002-2018-5714}}%
\author[a]{Samuel~J.~Clark%
  \IUCrEmaillink{sjclark@anl.gov}%
  \IUCrOrcidlink{0000-0002-8678-3020}}%
\author[a]{Francesco~De Carlo%
  \IUCrCemaillink{decarlo@anl.gov}%
  \IUCrOrcidlink{0000-0003-1068-7785}}%

\affil[a]{Advanced Photon Source, Argonne National Laboratory,
9700~South~Cass~Avenue, Lemont, IL~60439, USA}

\begin{document}
\maketitle

\begin{synopsis}
We develop and evaluate a deep learning-based algorithm to calibrate the rotation center of a synchrotron tomography instrument using tomogram slices of the target sample reconstructed with a range of center of rotation parameters. Our independent test results indicate the proposed method can achieve a sub-pixel mean absolute error with good model interpretability and maintain robust performance when the raw data are subject to reduction in the number of projections by a factor of 10 or Poisson's noise with a blank scan factor of 10.
\end{synopsis}

\begin{abstract}
Accurate determination of the rotation-axis position is a prerequisite for artifact-free reconstruction in parallel-beam synchrotron micro-tomography. Traditional approaches such as Vo's method rely on sinogram features that can fail for low-contrast or weakly absorbing specimens. We present a learning-based method that treats center selection as a binary classification problem, using a DINOv2-pretrained vision transformer aggregated with attention-based multiple-instance learning, fine-tuned end-to-end on tomographic images. At inference time, the proposed algorithm was applied to a stack of tomograms reconstructed at a sweep of candidate centers to select the optimal center for reconstruction. We tested the estimation accuracy of the proposed method on two independent data sources and consistently achieved a mean absolute error of below 1 pixel. We also tested the method robustness to sparse or noisy acquisitions with the same datasets and demonstrated consistent performance when the number of projections was reduced by a factor of up to 10 or the blank scan factor of the underlying Poisson's noise was increased to 10. We also illustrated the interpretability of the proposed method by mapping out the relative contributions of continuous spatial features to the overall classification task. This method, delivered as \emph{tomo-center}, an open-source command-line tool, has been integrated into several tomography software packages to assist experiments during the routine beamline operations.
\end{abstract}


\section{Introduction}

In parallel-beam tomography the center of rotation (COR) must be located to sub-pixel accuracy; a misalignment in the COR introduces characteristic ``double-edge'' and tangential streak artifacts that propagate through the entire reconstructed volume. Classical estimators include image-based registration of $0^{\circ}$/$180^{\circ}$ projection pairs~\cite{donath2006automated}, sinogram self-symmetry searches, and the Fourier-domain method of Vo \emph{et al.}~\cite{vo2014reliable}. These methods perform well on high-contrast samples with isotropic features, but degrade on weakly absorbing biological specimens, samples with strong local symmetry, and partial-field-of-view scans.\par

A well-established alternative to existing methods is via the use of a “try reconstruction”: a series of 2-D tomograms from one representative slice into the sample volume, reconstructed under a range of candidate COR parameters. More specifically, each tomogram from the try reconstruction is examined by the operator to manually determine the optimal COR that yields the highest reconstruction quality. The main disadvantage of the try-reconstruction method is its reliance on operator inspection, as well as the likely increase in reconstruction time. To overcome such disadvantages, neural networks (NN) have been investigated to automatically distinguish tomograms from the try reconstruction with the correct and incorrect COR parameters \cite{yang2017convolutional}. Accurate prediction results have been reported, as a proof of principle, from both synthetic and actual tomograms. To further bridge the technical gap between existing automated algorithms for COR detection and their widespread adoption in x-ray tomography experiments during the routine operations of synchrotron light sources, ongoing efforts to improve method generalizability, efficiency, and robustness are needed, as detailed below. \par

First, in recent years, various frameworks for deep neural network (DNN) pretraining \cite{caron2021emerging,zhou2021ibot,he2022masked} have unlocked great potential of existing unified model architectures to learn and extract general-purpose image features at scale. In particular, Caron and colleagues developed a self-supervised learning (SSL) framework named ``knowledge distillation with no labels'' (DINO) \cite{caron2021emerging} that extended knowledge distillation \cite{hinton2015distilling} and self-training \cite{xie2020self}. More specifically, in the DINO framework, the student model is trained using no labels, and guidance is only provided from a teacher model that is dynamically built during training. When applied to the vision transformer (ViT) model architecture, the DINO framework has demonstrated excellent performance on several benchmarks in both pretraining and finetuning settings. DINO v2 and v3 \cite{oquab2023dinov2,simeoni2025dinov3} have further scaled up the DINO training framework in terms of the data and model size, establishing a new state of the art particularly in in-domain zero-shot image understanding. Owing to the strong feature extraction capabilities of DNNs and the effectiveness of the DINO framework, models pretrained on large-scale, high-quality natural-image datasets transfer more effectively to downstream tasks in different domains, such as synchrotron x-ray tomography. \par

Second, tomography experiments at synchrotron light sources often acquire thousands of projection images and reconstruct slices containing several million pixels each \cite{liu2019deep}. The data volume scales with sample size, acquisition speed, and spatial resolution \cite{yang2017convolutional}, posing significant challenges for the efficiency of the underlying reconstruction pipelines. Moreover, given the complex structure of the samples, detector imperfections, and their interplay with the incident beam, the contributions of local image features to effectively distinguishing the correct COR parameter are often spatially heterogeneous. This setting is reminiscent of the multiple instance problem in artificial intelligence \cite{dietterich1997solving}, where each training example, termed a ``bag'', contains multiple instances, with supervision provided only at the bag level, and the objective is to learn a bag-level function, typically by aggregating representations of the instances within each bag. Multi-instance learning (MIL) was initially introduced to study problems such as drug activity prediction and molecular behavior analysis in mass spectrometry data, while its application to imaging challenges has later evolved in order to improve tasks such as anomaly detection in medical images and videos \cite{quellec2017multiple}. In particular, Ilse, Tomczak, and colleagues have proposed the use of neural networks in the MIL layout to classify, weigh individual instances, and combine instance-level responses into bag-level predictions in an end-to-end manner \cite{ilse2018attention}. This method has demonstrated competitive performance on MIL benchmarks while providing interpretability through the identification of key instances. More recent development in MIL has concentrated on modeling correlation among instances to improve classification accuracy, accelerate training convergence, and increase interpretability \cite{shao2021transmil}. \par

Third, when the central axis of the sample stage is significantly misaligned with the central axis of the detector, which could be deliberate for double FOV scans, or dynamic sample environments are used which may cause the COR to wander, the capture range of the COR estimation for reconstruction must be both robust and wide. \par

In this work we investigate a learning-based alternative: a stack of slices reconstructed at a sweep of candidate centers is presented to a DINOv2~\cite{oquab2023dinov2} vision transformer with attention pooling and a binary classification head trained under the MIL framework to discriminate ``correctly aligned'' from ``misaligned'' reconstructions, and the candidate yielding the highest softmax score is returned as the estimated rotation center. Multiple random windows drawn from each slice are aggregated with attention-based MIL~\cite{ilse2018attention} to provide robustness to local sample heterogeneity. We deliver the method as \emph{tomo-center}~\cite{tomocenter2026}, a stand-alone command-line package. The remainder of the paper describes the model architecture, its training, validation (Section~\ref{sec:method}), testing procedures (Section~\ref{sec:comparison}), qualitative and quantitative results on representative synchrotron datasets (Section~\ref{sec:results}), and example applications that integrate the classifier into the \emph{tomocupy} and \emph{tomo-gui} reconstruction package and interface as an end-to-end sweep-and-reconstruct workflow (Section~\ref{sec:application}). \par

\section{Method}
\label{sec:method}
\subsection{Model architecture}
\begin{figure}[ht]
\begin{center}
 \includegraphics[width=1\textwidth]{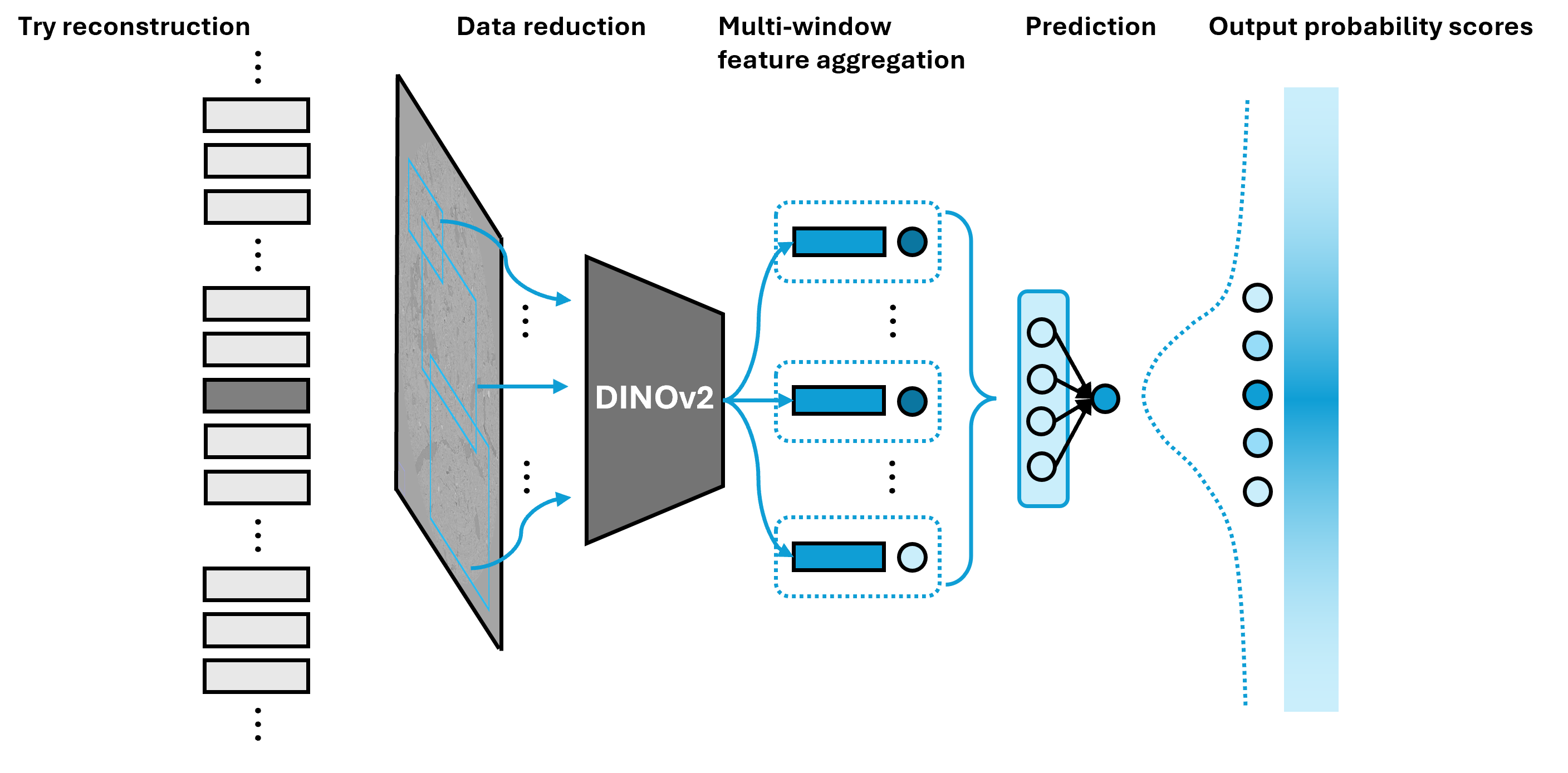}
\end{center}
\caption{Schematic of the proposed framework to predict the correct COR based on the corresponding try reconstruction.}
\label{fig:model_schematic}
\end{figure}
Figure \ref{fig:model_schematic} illustrates the proposed framework to classify the underlying COR parameter of each tomogram from the try reconstruction as correct and incorrect. More specifically, $K$ windows, each of size $s\times s$, are uniformly sampled from each input tomogram and fed to the ViT-B/14 backbone for feature extraction. The resulting instance-level class tokens are aggregated into a single bag-level token through attention pooling \cite{ilse2018attention}, with the attention weight associated with the $k^{th}$ window ($k=1,...,K$) computed following the gated attention mechanism:
\begin{equation}
a_k \;=\; \frac{\exp(\mathbf{w}^T (tanh(\mathbf{V} \mathbf{h}_k^T)\odot sigm(\mathbf{U} \mathbf{h}_k^T)))}{\sum_{j=1}^{K} \exp(\mathbf{w}^T (tanh(\mathbf{V} \mathbf{h}_j^T)\odot sigm(\mathbf{U} \mathbf{h}_j^T)))},
\label{eq:gated_attention_weights}
\end{equation}
where $\mathbf{h}_j$ ($j=1,...,K$) is the class token corresponding to the $j^{th}$ window and $\mathbf{w}$, $\mathbf{V}$, and $\mathbf{U}$ are trainable model weights. The pooled token is then projected into two-class logits $f_0$ and $f_1$ by a linear classification head and transformed into probabilities $p_0$ and $p_1$ by the softmax operator.\par  

\subsection{Model training}
\subsubsection{Data}
\label{sec:training_data}
The tomographic training dataset was compiled from a total of 546 retrospective experiments covering a wide range of samples including bone, wood, and soil. For each experiment, the try reconstructions were achieved using both the center and off-center slices (i.e., slices perpendicular to the rotation axis) by sweeping the COR parameter over a range of up to 490 steps in 0.5-pixel increments. In particular, the median value and inter quartile range (IQR) of the number of steps were both 200. Apart from the COR parameter, other reconstruction parameters were either set to the user-validated optimal values or left at their default settings, resulting in around 110,000 tomograms for model finetuning.\par
\subsubsection{Pretraining}
\label{sec:pretrain}
Weights of the ViT-B/14 backbone were initialized from the checkpoint released by Meta following the DINOv2 framework \cite{oquab2023dinov2}. Weights of the attention pooling layers were trained from scratch. \par
\subsubsection{Finetuning}
\label{sec:finetune}
For each tomogram during model finetuning, the min-max normalization was first applied. A random horizontal flip was applied for data augmentation to generate tomograms when the sample was mounted in an inverted orientation. A total of 24 windows of size $518\times518$ were then randomly sampled from each tomogram to generate local patches. To further desensitize the model to variations in the region of interest (ROI), orientation, and scale of the image features, a second square window of the length uniformly sampled from 66\%, 75\%, 87.5\%, and 100\% of that of the first window, and with corner coordinates uniformly sampled from a fixed set of 13 offsets was used to crop each local patch, which was subsequently resized back. Last, a rotation uniformly sampled from multiples of 90 degrees was applied to each resampled patch. The training loss was defined as the cross-entropy loss between the binary label of each tomogram indicative of the correctness of the underlying COR and the model output, as follows:
\begin{equation}
l(\mathbf{x},\mathbf{y})=\frac{1}{N}\sum_{n=1}^{N}(-y_{n,c}\frac{e^{x_{n,1}}}{e^{x_{n,0}}+e^{x_{n,1}}}),
\label{eq:ce_loss}
\end{equation}
where $x_{n,c}$ and $y_{n,c}$ are the logits and class labels for class c and sample n in the batch. For binary classification, $y_{n,c}$ simplifies to $y_n$ as the correctness of the underlying COR for sample n. In this work, a tolerance of $\pm 1$ pixel centered at the expert-annotated COR was applied to create the labels for each try reconstruction in the training data.\par

Model training was performed using the PyTorch framework on the Polaris HPC system at the Argonne
Leadership Computing Facility (ALCF). Data parallelization was implemented by initiating multiple processes across 4 Nvidia A-100 SXM4 GPUs (40GB memory each) on each computing node for a total of 20 nodes. The model was trained for a total of 10 epochs. During each epoch, the training set was enlarged by a factor of 6 through bootstrap sampling. Since the correct and incorrect CORs in each series of try reconstruction were highly imbalanced, tomograms with the incorrect COR were further down-sampled. These samples were distributed uniformly across all processes. Within each process, the batch size was set to 2. The AdamW optimizer was used to minimize the training loss, and the initial learning rate was set to $5\times 10^{-6}$ and updated following a cosine annealing scheduler. The first 500 training iterations were used for warmup (i.e., the initial learning rate was scaled by the fractional iteration number in this warmup window).\par

Validation was performed after every training epoch (i.e., 205 iterations) using randomly selected try reconstructions from tomography experiments of the bone and soil samples respectively. A total of 20 independent experiments, 10 from each category were included. For each try reconstruction, the COR parameters covered a range of 200 steps in 0.5-pixel increments. In terms of other reconstruction parameters, they were set to user-validated optimal values for the try reconstructions of the bone samples and left at their default settings for the try reconstructions of the soil samples, respectively. The tomogram image sizes of the bone and soil samples were $2448\times 2448$ and $3232\times 3232$. Similar to the preprocessing of the training data, min-max normalization was applied to the input images. A total of 3 windows of size $518\times 518$ were then randomly sampled from each tomogram to generate local patches, which were subsequently input to the proposed model to predict the probability score that the underlying tomogram was reconstructed under the correct COR. Among all the CORs used to generate each try reconstruction, the one corresponding to the tomogram with the highest probability score was determined as the model-predicted COR and its absolute error relative to the expert annotation was evaluated. For each validation time point, mean absolute errors (MAEs) across all 20 try reconstructions were further averaged. \par

\subsection{Usage}
The algorithm
is implemented in Python and publicly available at \url{https://github.com/xray-imaging/tomo-center}. Once installed, \emph{tomo-center}~\cite{tomocenter2026} is invoked through two subcommands,
\verb|tomo-center find| (inference) and \verb|tomo-center train|
(fine-tuning). Full command-line
reference, including all flags, default values, input-folder
conventions, and worked end-to-end examples for both subcommands, is
maintained in the package README and kept in sync with
each release. \par

\section{Performance evaluation}
\label{sec:comparison}

In this section, we
benchmark the proposed method on 1) a production-scale dataset and 2) eight selected datasets from the TomoBank repository \cite{de2018tomobank} that are fully
disjoint from the training set described in Section~\ref{sec:training_data}. For the first data source, we
compare the proposed method exposed by
\emph{tomocupy}~\cite{nikitin2023tomocupy} as
\texttt{--rotation-axis-method ai} against
the SIFT-based method~\cite{lowe2004sift}
(\texttt{--rotation-axis-method sift}). For the second data source, we compare the proposed method against both Vo's
method~\cite{vo2014reliable} (\texttt{--rotation-axis-method vo}) and the SIFT-based method. All three configurations operate
on the same input scans. Results are reported in Section \ref{sec:quantitative}. \par

\subsection{An in-situ dataset designed as a robustness benchmark}
\label{sec:comp_dataset}

The benchmark comprises 305 micro-tomography scans collected at beamline
2-BM of the Advanced Photon Source (APS) during an \emph{in-situ} study of
calcium carbonate carbonation. The dataset is dominated by two thermally
programmed experiment sequences (104 and 190 evaluated scans
respectively) in which a single specimen was repeatedly scanned while
the sample environment temperature was ramped between 20\,$^{\circ}$C
and 800\,$^{\circ}$C, plus a handful of single-acquisition reference
scans. Every acquisition is a 180\textdegree\ sweep with 1\,501
projections. The two experiment sequences used two different sensor
ROIs: 2\,048\,$\times$\,2\,048\,pixels for the first sequence and
2\,448\,$\times$\,2\,448\,pixels for the second.
Critically, none of the specimens or their composition were present in
the data used to train the proposed model (Section~\ref{sec:method}); the
dataset is therefore an out-of-distribution test of the shipped model. \par

Three properties of the experimental design make this dataset
particularly well suited to a COR robustness comparison:
\begin{itemize}
  \item \textbf{Fixed rotation axis.} The thermal stage was assembled
        once at the start of each sequence, the rotation axis was
        mechanically aligned close to the geometric centre of the
        detector, and neither the stage nor the camera ROI was
        repositioned for the duration of the experiment. The rotation
        axis was therefore fixed across both measurement sequences and
        held within a few pixels of $n_x/2$ throughout the campaign.
  \item \textbf{Common flat-field reference.} The flat (white) fields
        were acquired once, at the beginning of each sequence and with
        the furnace already in place, so every scan in the sequence
        shares the same intensity normalisation.
  \item \textbf{Strong sample evolution.} The specimen itself changed
        substantially as the temperature ramp proceeded: pore morphology,
        attenuation contrast and local symmetry all evolved with the
        carbonation reaction. The rotation axis is therefore the only
        quantity in the system held intentionally constant across the
        thermal program.
\end{itemize}
Together these properties give a clean test of method robustness: any
per-scan COR error must be attributed to the COR estimation algorithm
rather than to a moving target. The relevant statistic across the
sequence is the dispersion of each method's prediction around the
fixed mechanical alignment. \par

We adopt Vo's per-scan output as the ground truth for this benchmark:
Vo's values were used to produce the production reconstructions
delivered to the user and were operator-validated by inspecting the
reconstructed volumes against the mechanically-aligned axis. Vo's per-scan values
coincided with the geometric image centre $n_x/2$ to within $\sim 5$
pixels for every scan in this campaign. One scan was excluded from
the analysis because the sample shifted physically during a single
acquisition, and two further scans had no entry in the production
reconstruction script (no SIFT recipe available). The effective
benchmark size shared across all three configurations is $n=302$. For the proposed method and the Vo's method, a range of 200 steps in 0.5-pixel increments was used consistently.\par

\subsection{Eight datasets from the TomoBank repository}
\label{sec:tomobank_dataset}
Another 8 independent cases from the TomoBank repository \cite{de2018tomobank} (with data and meta data available\footnote{\url{https://tomobank.readthedocs.io/en/latest/source/data/docs.data.roundrobin.html}}\footnote{\url{https://tomobank.readthedocs.io/en/latest/source/data/docs.data.dynamic.html\#methane-hydrate-formation}}) were used as the testing data. COR estimates from the proposed method, the Vo's method and the SIFT method were all compared with expert-annotated ground truth. For the proposed method and the Vo's method, a range of 400 steps in 0.5-pixel increments was used consistently.\par

\subsection{Stress tests: sparse-angle and low-photon acquisitions}
Beyond the unperturbed evaluation of the preceding sections, we
performed two stress tests to assess COR estimation robustness under
acquisition compromises that are routinely made at beamlines: reducing
the number of projections, or reducing the exposure time per
projection. Both compromises are made to shorten the total scan time
when fast sample evolution must be captured within a single scan, when
the available beamtime is limited, or to reduce the integrated radiation
dose on dose-sensitive specimens. We ran both scenarios on the same 10
randomly-selected scans of the in-situ benchmark in Section \ref{sec:comp_dataset} with the
unperturbed-data Vo COR taken as the per-scan ground truth, and all applicable scans selected from the TomoBank repository in Section \ref{sec:tomobank_dataset}, respectively. Since the SIFT method only relies on two projections with angles $180^\circ$ apart, and its underlying key point detection may require further optimization in the hyper-parameters, it was excluded from the stress tests. \par

\textbf{Sparse-angle datasets.} Three sparsity levels were tested,
keeping every 5\textsuperscript{th}, every 10\textsuperscript{th}, and
every 20\textsuperscript{th} projection — yielding 300, 150, and 75 of
the original 1\,501 projections per scan for the in-situ benchmark in Section \ref{sec:comp_dataset}. For the datasets selected from the TomoBank repository, the original numbers of projections ranged from 900 to 1500 and the same 3 down-sampling factors were used. For each sample and each
level, the raw HDF5 was rewritten with only the selected projections
and the corresponding theta values; the reconstruction recipe (Paganin
phase retrieval, Fourier-wavelet stripe removal, Shepp--Logan FBP
filter) was otherwise unchanged. \par

\textbf{Low-photon datasets.} Three photon-scaling levels were tested:
$1/10$, $1/100$ and $1/1000$ of the original photon flux. For each
sample and each level, the raw projection intensities $P_n$ were re-sampled from $Poisson(\frac{P_n}{N})$ and scaled by $N$, where $N$ in {10, 100, 1000} is the
photon-reduction factor (blank scan factor). This preserves the mean intensity while
increasing the per-pixel noise standard deviation by $\sqrt{N}$
relative to baseline; this preserves the mean intensity
while injecting noise consistent with the reduced photon count. Similar to the sparse-angle dataset, the raw HDF5 was rewritten with the noise injected into the projection data, whereas the
projection count and reconstruction recipe were unchanged. In particular, one dataset presented in Section \ref{sec:tomobank_dataset} was excluded from the stress test for low-photon flux due to its intrinsic high noise level. \par

\subsection{Methods compared}
\label{sec:comp_methods}

We evaluated three centre-finding configurations:
\begin{itemize}
  \item \textbf{Proposed method} -- the \emph{tomo-center} checkpoint finetuned and validated according to Section \ref{sec:finetune}. Hyper-parameters at the inference time refer to those used for model validation in Section \ref{sec:finetune}.
  \item \textbf{Vo} -- the Vo method~\cite{vo2014reliable} exposed by
        \emph{tomocupy} as \texttt{--rotation-axis-method vo}, which
        searches the rotation axis in sinogram space by minimising a
        sinogram self-symmetry metric. For the first data source, this method is used as the ground truth as validated by the operator.
  \item \textbf{SIFT} -- the classical SIFT centre-finder exposed by
        \emph{tomocupy} as \texttt{--rotation-axis-method sift}, which
        detects keypoints on the 0\textdegree/180\textdegree\
        projection pair using OpenCV's SIFT
        implementation~\cite{lowe2004sift} and derives the centre from
        the mean horizontal shift between matched keypoints.
\end{itemize}
All three configurations were run on the same single NVIDIA GPU at the
beamline workstation, with the same try-centre inputs (for the AI
configurations) and the same per-scan dark/flat references (for Vo and
SIFT).\par

\section{Results}
\label{sec:results}

\subsection{Training curves}
\begin{figure}[ht]
\begin{center}
 \includegraphics[width=1\textwidth]{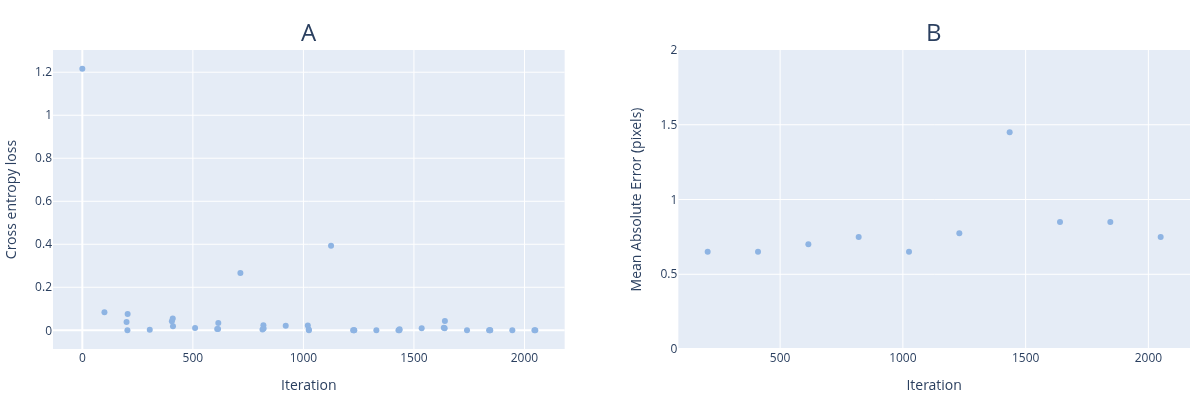}
\end{center}
\caption{Training losses (A) and validation errors (B) of the proposed model to classify the COR correctness given the underlying try reconstructions during fine tuning.}
\label{fig:training_curve}
\end{figure}

Figure \ref{fig:training_curve} shows the training losses and validation errors of the proposed model for COR detection during model fine tuning. Overall, a converging trend can be observed. At the validation time, the MAE attains the lowest value within the first training epoch, and shows a slight increasing trend afterwards. As a result, the model weights saved at the end of the first training epoch were used. \par

\subsection{Qualitative analysis}
\begin{figure}[ht]
\begin{center}
\includegraphics[width=\textwidth]{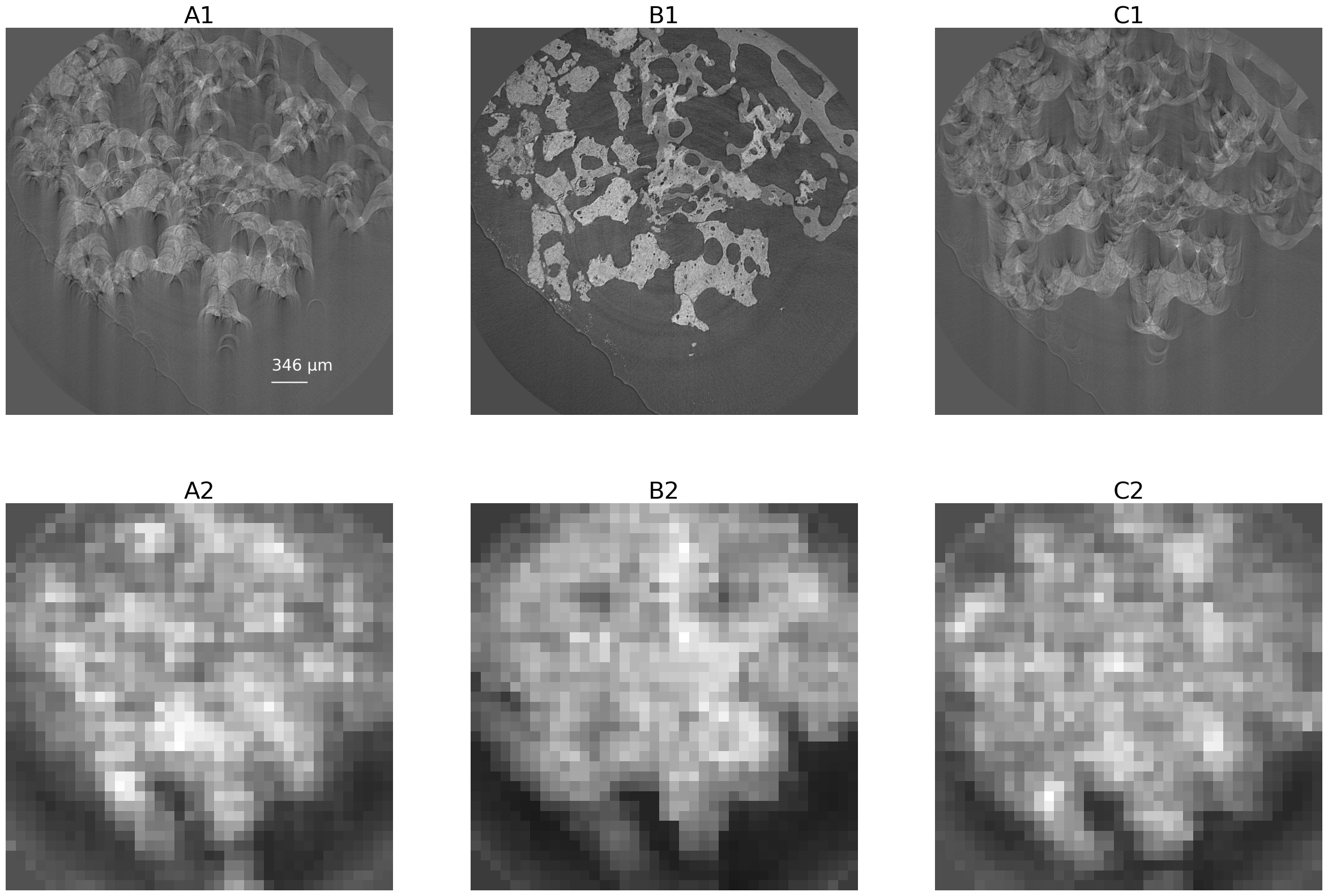}
\end{center}
\caption{Selected tomograms (row 1) and their corresponding distributions of spatial importance (row 2) as learned and utilized by the proposed model to predict the correctness of the underlying COR reconstruction parameter. Columns A-C represent the underlying CORs that are 50 pixels less than, identical to, and 50 pixels more than the actual COR, respectively. The predicted probabilities of correctness are 0.00, 0.99, and 0.01, respectively.}
\label{fig:attention_maps}
\end{figure}

Figure \ref{fig:attention_maps} shows an exemplar set of the validation data (row 1) and the spatial importance learned by the proposed model through attention pooling (row 2) to determine the correctness of the underlying COR. When the underlying COR parameters significantly deviate from the actual parameter, the spatial organization of the sample is severely distorted in the reconstructed tomograms (A1 and C1), as compared to the same reconstructed with the correct parameter (B1), consistently with previous studies \cite{yang2017convolutional}.\par
In order to identify the key instances to the bag-level model prediction as a standard test for MIL interpretability \cite{ilse2018attention}, contributions of distinct patches that densely covered the entire tomogram were quantified and organized in the form of an attention map. More specifically, the patch size was set to $224\times 224$ with a stride size of $56\times 56$, resulting in a total of 1600 image patches across the ROI that were jointly input to the model. Both the final logits for whole-image classification and the weights for all patches computed during the gated attention stage were extracted. According to row 2 of Figure \ref{fig:attention_maps}, the model learned to consistently use the foreground region (and ignore the background region) of the tomogram to aggregate evidence for both the correct and incorrect underlying COR parameters and achieved accurate classification results. Similar results have been observed from another 3 distinct cases from the validation set as shown in the supplementary material (Figures S1-S3).\par
Figure~\ref{fig:stress_test_comparison} compares two reconstructed tomographic slices (one from each data source in Section \ref{sec:comparison}) using the COR predicted by the proposed method and by the Vo's method when the raw data are subject to Poisson’s noise of blank scan factor of 100. As can be seen from the first two columns, the image quality decreases significantly due to the presence of the Poisson's noise, creating challenging conditions for the normal operation of the reconstruction pipeline. When the proposed method is compared with Vo's method, the proposed method consistently predicts more accurate COR parameters. As is shown in columns C and D, the tomogram slices reconstructed with CORs predicted by the proposed method closely match those reconstructed with the ground truth under the same conditions, whereas the Vo-derived centers produce the double-edge and tangential streak artifacts characteristic of a misaligned axis. \par

\subsection{Quantitative analysis}
\label{sec:quantitative}
\subsubsection{The first data source}
\paragraph{Accuracy and running time}
\label{sec:comp_headline}

Table~\ref{tab:comp_headline} reports the absolute deviations of the CORs estimated using
the proposed method and the SIFT method from those estimated using the Vo's method (the production reference)
on the 302 evaluated scans. Both methods agree with Vo to within
sub-pixel accuracy: the proposed method recovers Vo's centre to
within one pixel on 301 of the 302 scans and to within two pixels
on all 302; SIFT agrees with Vo within two pixels on every scan.
The three methods are essentially equivalent on this dataset. In addition, the agreement holds across both experiment sequences
(Table~\ref{tab:comp_size}): on the smaller-crop sequence, the
proposed method's mean deviation from Vo is 0.30\,pixels and SIFT's
is 0.52\,pixels; on the larger-crop sequence, the values are
0.32\,pixels and 0.47\,pixels, respectively. Vo runs at $\sim 1$\,s
per scan and so is by far the fastest method; the proposed method
and SIFT are each $\sim 25$--30\,s per scan. \par

\begin{table}[ht]
\caption{Absolute deviation from the Vo reference on the
302-scan out-of-distribution benchmark. Errors are in detector
pixels. Running time is the per-scan wall-clock time including I/O.
Vo is the reference and is omitted from the table (deviation zero
by construction).}
\smallskip
\begin{center}
\begin{tabular}{lrrrrrrrr}
\midrule
Configuration & $N$ & mean & median & max & $\leq$0.5\,px & $\leq$1\,px & $\leq$2\,px & time (s) \\
\midrule
Proposed     & 302 & 0.32 & 0.50 & 2.00 & 277/302 & 301/302 & 302/302 & 24.0 \\
SIFT                    & 302 & 0.48 & 0.46 & 1.62 & 177/302 & 297/302 & 302/302 & 28.9 \\
\midrule
\end{tabular}
\end{center}
\label{tab:comp_headline}
\end{table}

\begin{table}[ht]
\caption{Per-family mean absolute deviation from the Vo reference.
The two in-situ sequences each contain many repeated scans on a
single specimen at varying temperature.}
\smallskip
\begin{center}
\begin{tabular}{lrrrrrr}
\midrule
Family & $N$ & Proposed & SIFT \\
\midrule
In-situ sequence A (2\,048$\times$2\,048) & 104 & 0.30 & 0.52 \\
In-situ sequence B (2\,448$\times$2\,448) & 190 & 0.32 & 0.47 \\
Single-acquisition refs                    &   8 & 0.38 & 0.43 \\
\midrule
\end{tabular}
\end{center}
\label{tab:comp_size}
\end{table}

\paragraph{Stability across the temperature ramp}
\label{sec:comp_temperature}

Because the rotation axis was mechanically fixed while the specimen
evolved substantially through the thermal program, the per-temperature
error distribution probes method robustness to sample-state changes
directly. Table~\ref{tab:comp_temperature} shows the per-temperature
mean absolute deviation from Vo for the proposed method and for
SIFT, on the two in-situ sequences. Errors of both methods remain sub-pixel
from 20 to 800\,$^{\circ}$C with bin-to-bin variation of less than
0.5\,pixels, demonstrating that the COR predictions are robust to
the morphological evolution of the carbonating specimen and that
neither method drifts from the Vo reference across the thermal
program. \par

\begin{table}[ht]
\caption{Mean absolute deviation from the Vo reference binned by
Eurotherm set-point, for the two in-situ sequences. Both the
proposed method and SIFT remain sub-pixel across the full
20--800\,$^{\circ}$C range.}
\smallskip
\begin{center}
\begin{tabular}{rrrrrrr}
\midrule
        & & \multicolumn{2}{c}{Sequence A} & &\multicolumn{2}{c}{Sequence B} \\
set-point (\,$^{\circ}$C) & $N_A$ & Proposed & SIFT & $N_B$ & Proposed & SIFT \\
\midrule
 20 & 17 & 0.26 & 0.57 & 12 & 0.08 & 0.59 \\
 25 &  - &   -- &  --  &  2 & 0.25 & 0.59 \\
450 &  - &   -- &  --  & 13 & 0.15 & 0.58 \\
600 & 39 & 0.29 & 0.43 &  7 & 0.29 & 0.42 \\
650 &  - &   -- &  --  & 19 & 0.24 & 0.47 \\
700 & 15 & 0.57 & 0.64 & 36 & 0.35 & 0.41 \\
800 & 33 & 0.21 & 0.54 &100 & 0.39 & 0.46 \\
\midrule
\end{tabular}
\end{center}
\label{tab:comp_temperature}
\end{table}

\paragraph{Stress tests}
\label{sec:comp_stress_test}
Absolute errors aggregated across
the 10 samples are reported in Table~\ref{tab:stress_test_insitu}. It is clear that both methods remain accurate when the number of projections during a tomographic scan is reduced by a factor of up to 10, with the median and IQR of the absolute errors at the sub-pixel level. In comparison, when only every $20^{th}$ projection from the original scan is available, both methods show more pronounced absolute errors. Overall, the proposed method and Vo's method show similar absolute errors with different reduction factors in the number of tomographic projections. In terms of the blank scan factor, both methods show consistently low errors when the blank scan factor is 10 and an increasing trend in the absolute errors with the blank scan factor. The proposed method is also markedly more robust to the underlying Poisson's noise. In particular, when the blank scan factor is 100, the proposed method maintains a sub-pixel accuracy in the median and IQR of the COR absolute errors, whereas the Vo's method results in a median of 8.25 pixels and an IQR of 39.25 pixels of the absolute errors. \par

\begin{figure}[ht]
\begin{center}
\begin{minipage}{0.24\textwidth}
  \centering
  \small A1\\[2pt]
  \includegraphics[width=\textwidth]{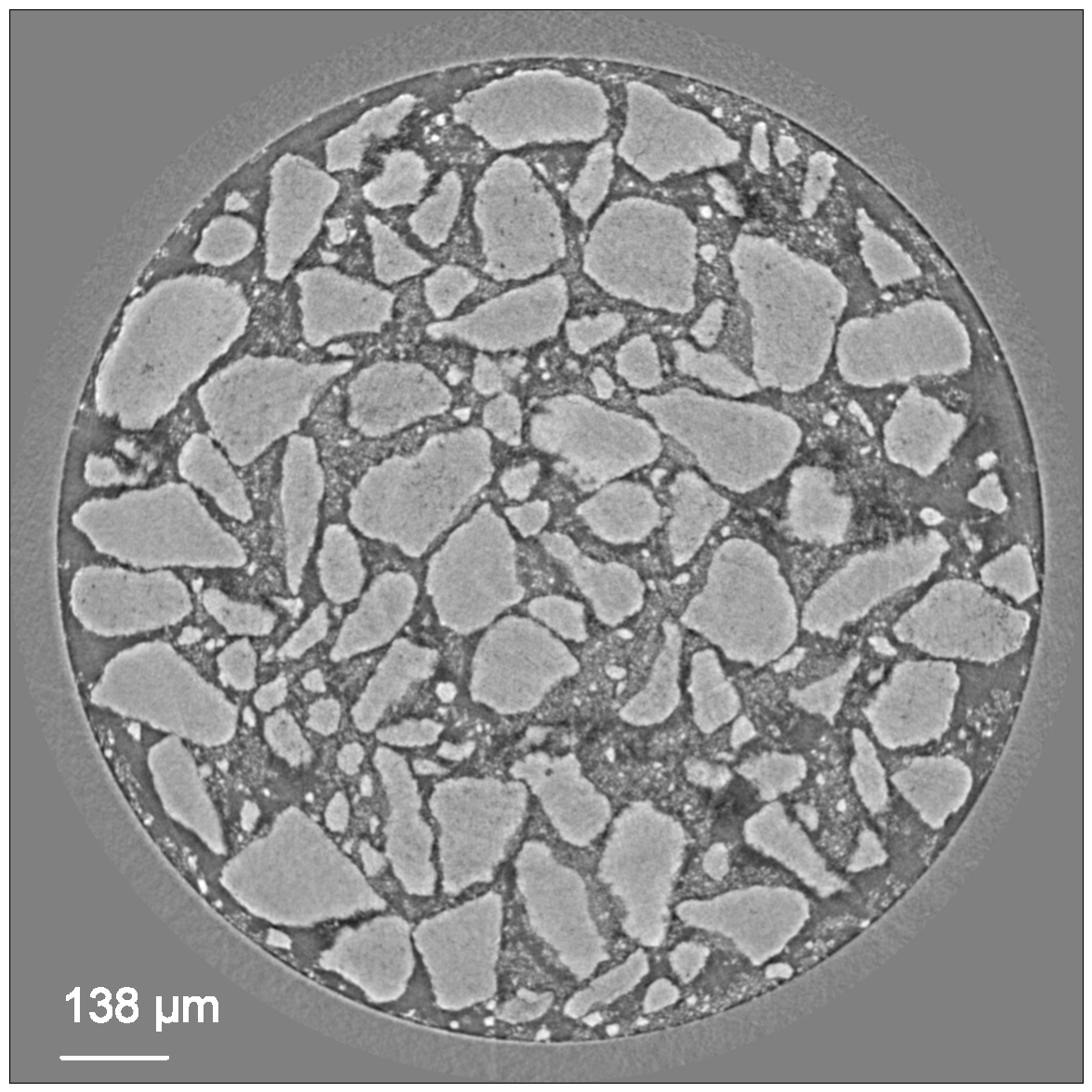}
  
\end{minipage}\hfill
\begin{minipage}{0.24\textwidth}
  \centering
  \small B1\\[2pt]
  \includegraphics[width=\textwidth]{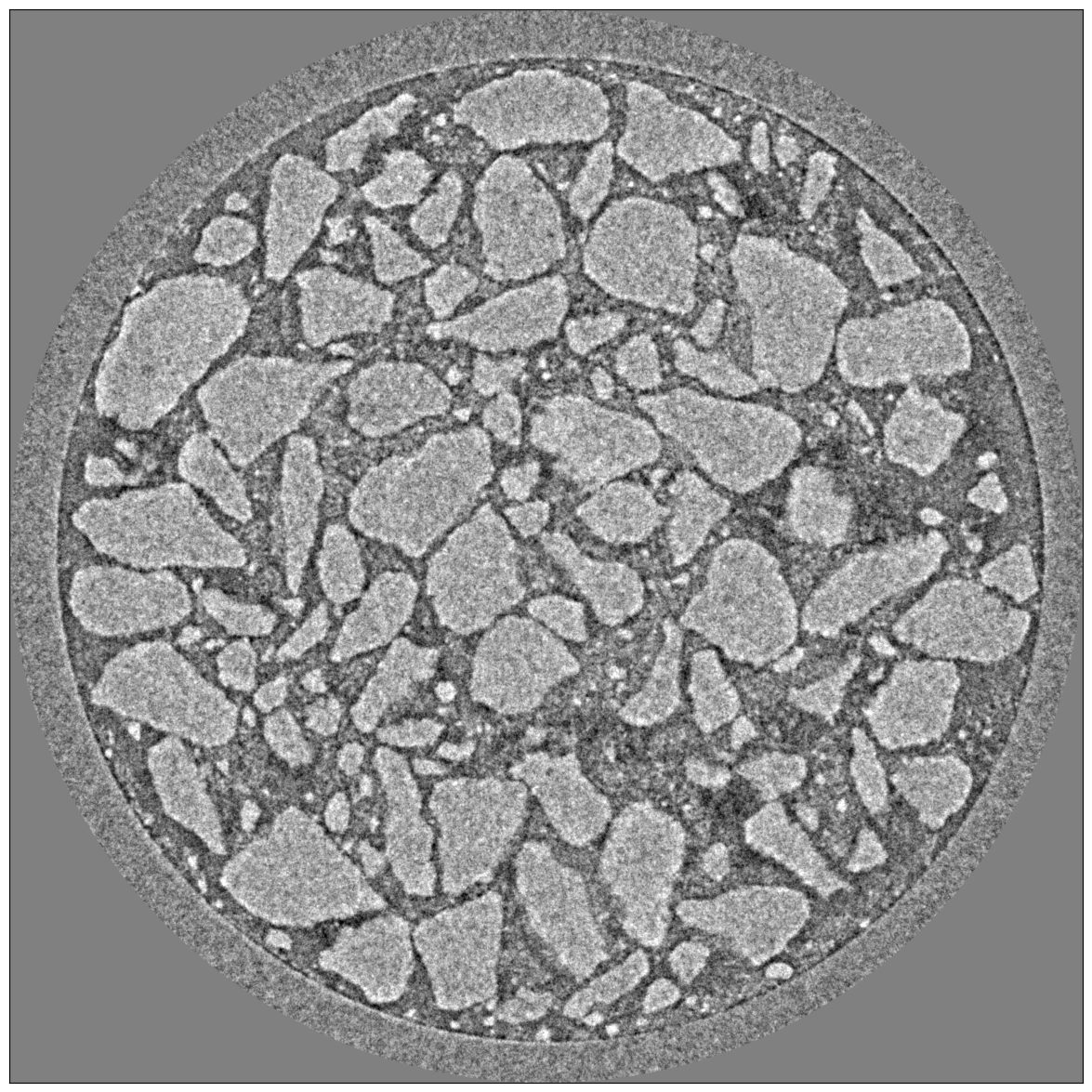}
\end{minipage}\hfill
\begin{minipage}{0.24\textwidth}
  \centering
  \small C1\\[2pt]
  \includegraphics[width=\textwidth]{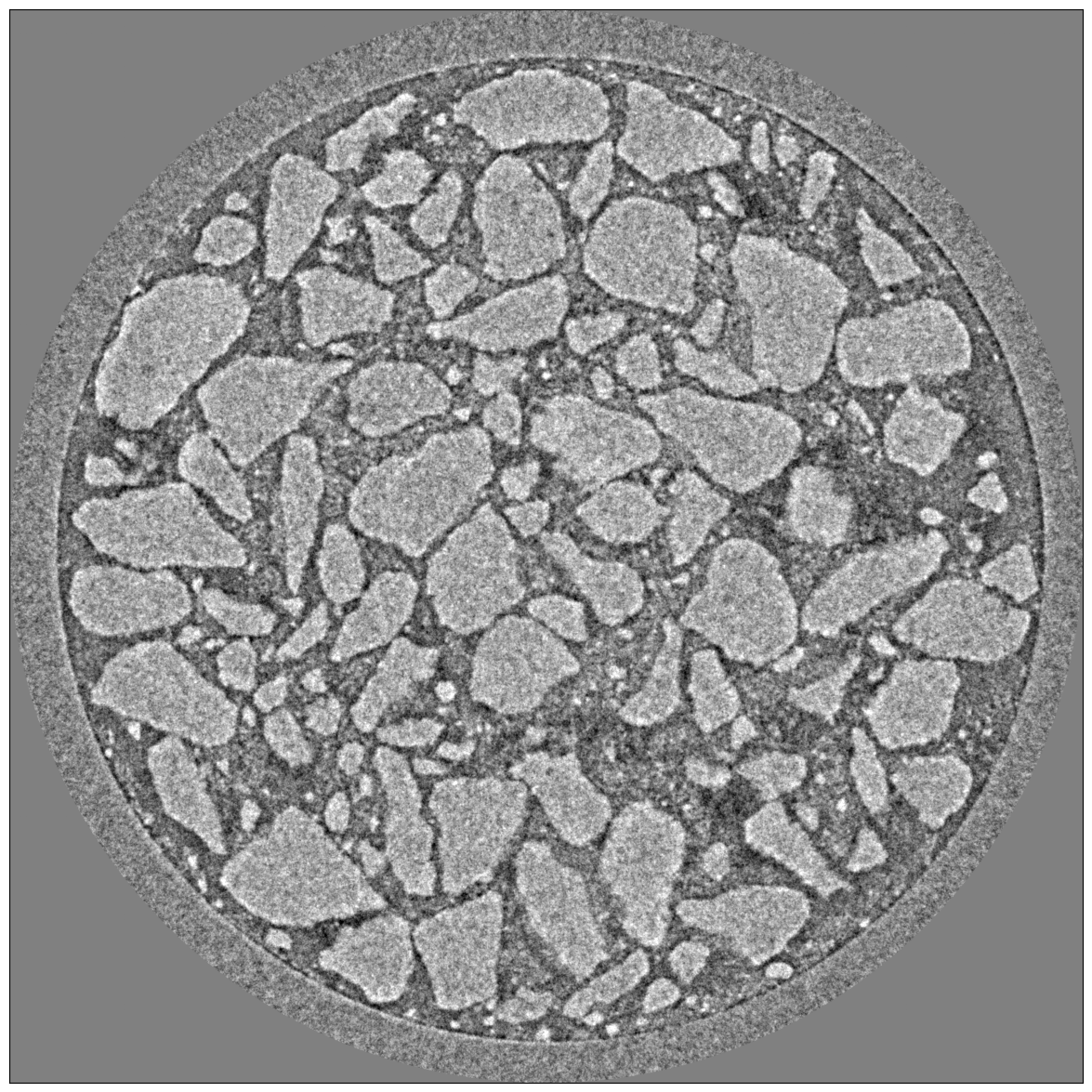}
\end{minipage}
\begin{minipage}{0.24\textwidth}
  \centering
  \small D1\\[2pt]
  \includegraphics[width=\textwidth]{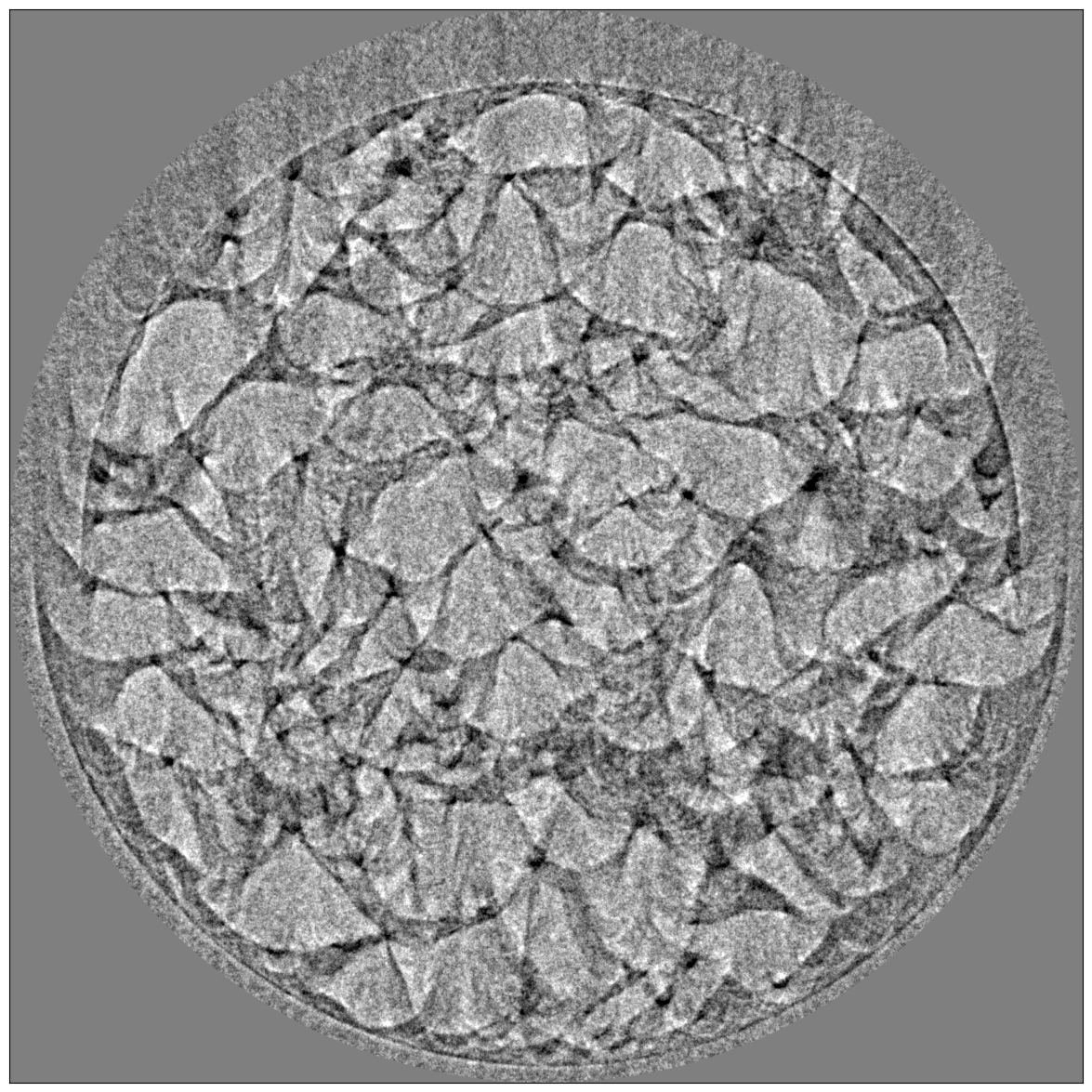}
\end{minipage}\\
\begin{minipage}{0.24\textwidth}
  \centering
  \small A2\\[2pt]
  \includegraphics[width=\textwidth]{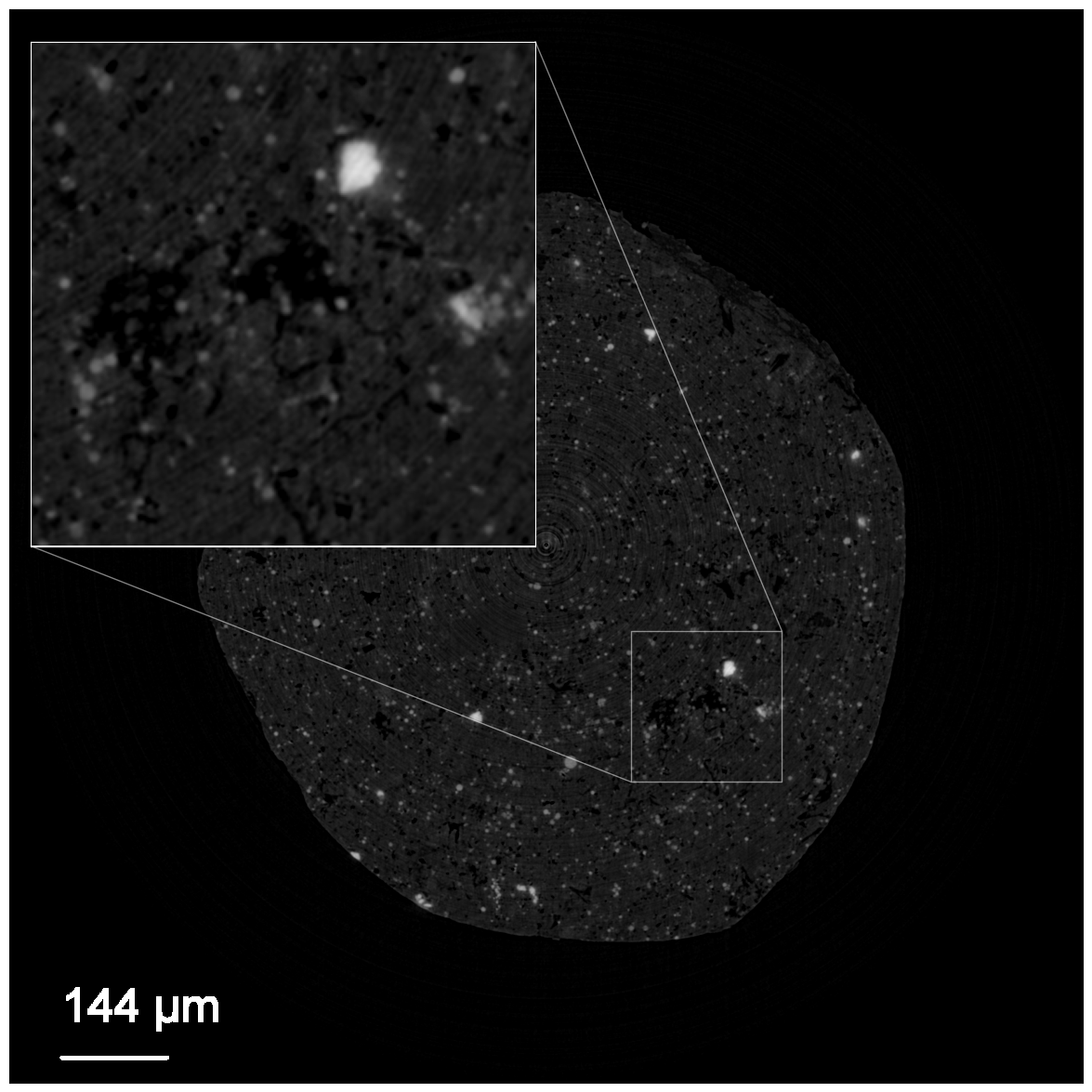}
\end{minipage}\hfill
\begin{minipage}{0.24\textwidth}
  \centering
  \small B2\\[2pt]
  \includegraphics[width=\textwidth]{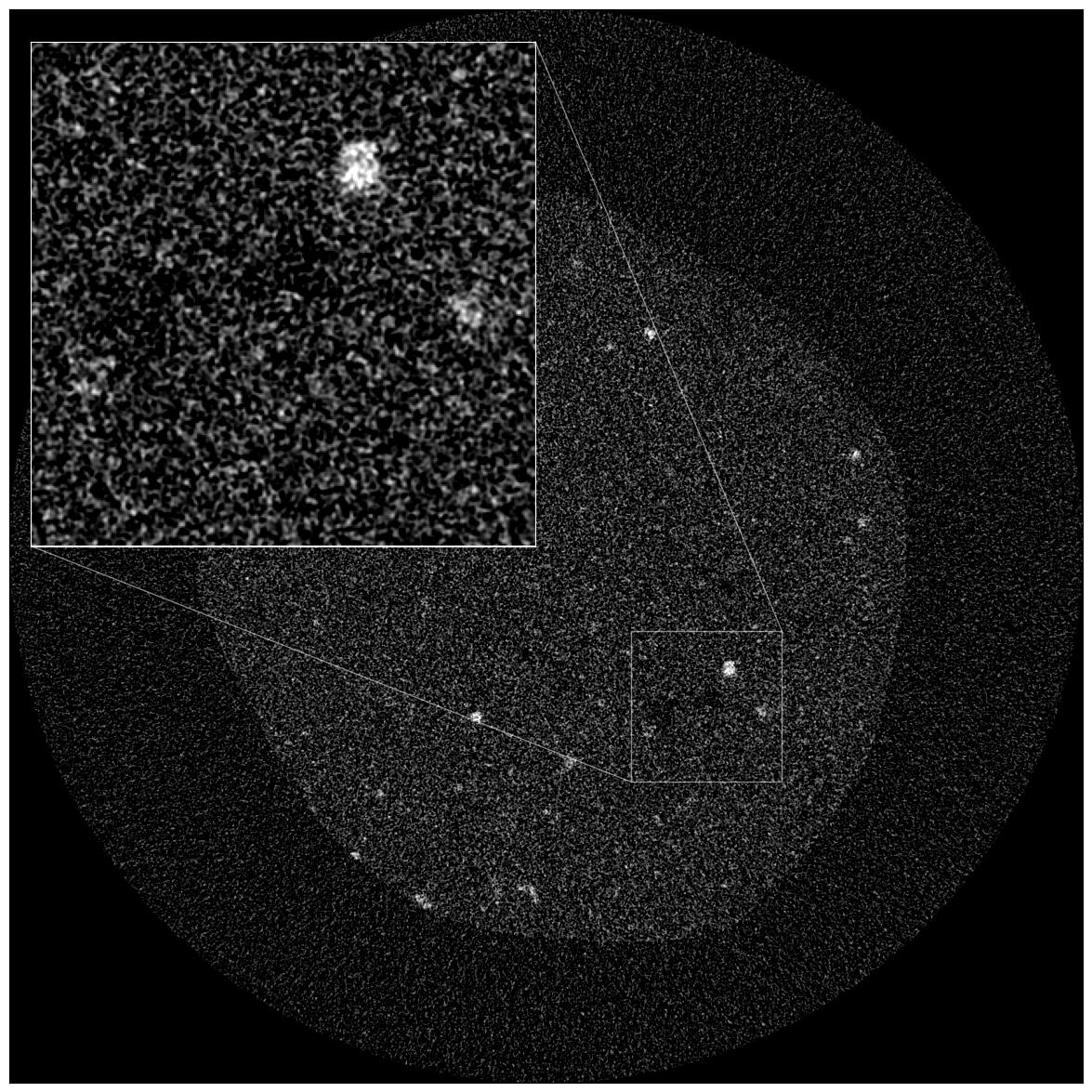}
\end{minipage}\hfill
\begin{minipage}{0.24\textwidth}
  \centering
  \small C2\\[2pt]
  \includegraphics[width=\textwidth]{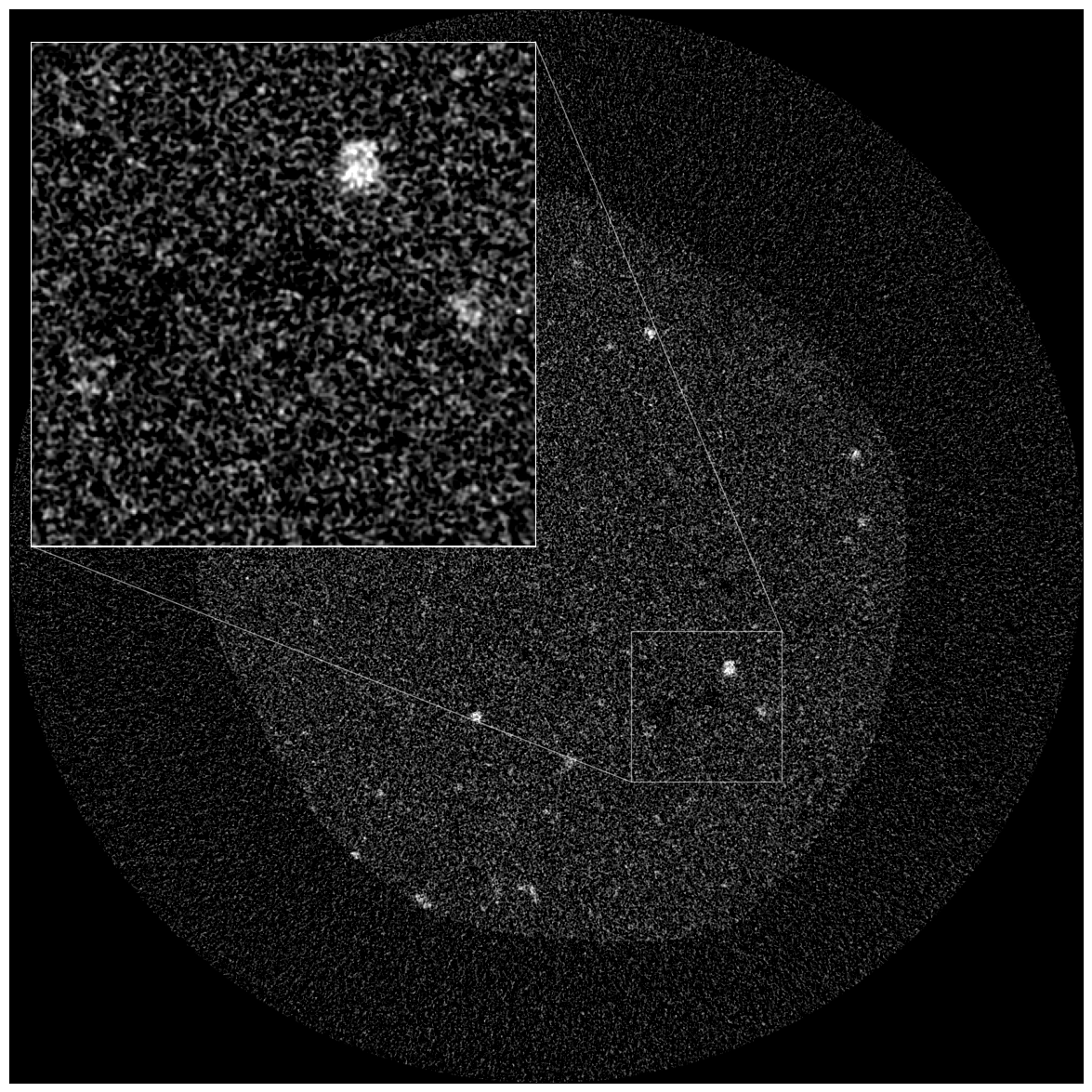}
\end{minipage}
\begin{minipage}{0.24\textwidth}
  \centering
  \small D2\\[2pt]
  \includegraphics[width=\textwidth]{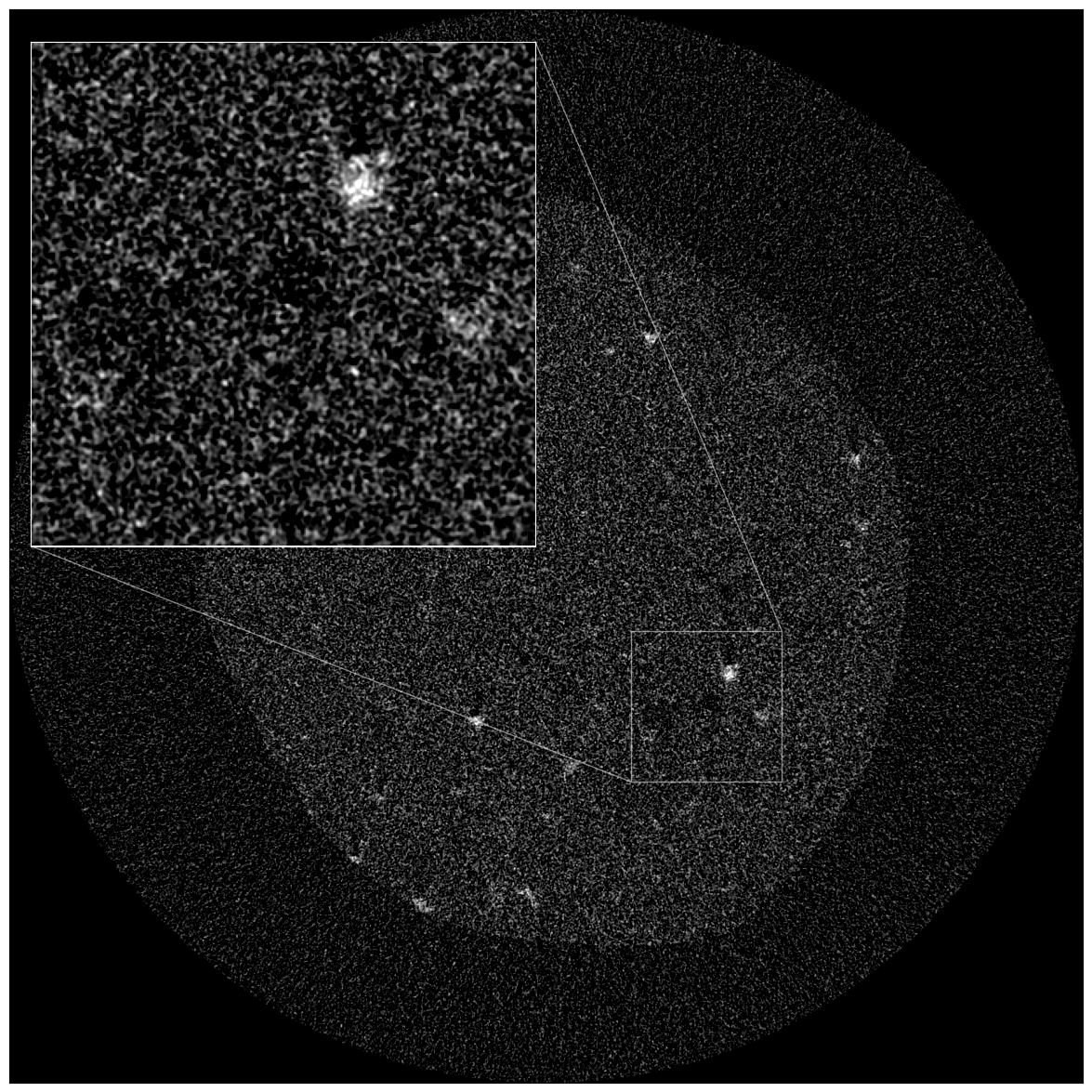}
\end{minipage}
\end{center}
\caption{Selected tomographic slices reconstructed with the correct COR (columns A and B), and CORs predicted by the proposed method (column C) and the Vo's method (column D) from an in-situ calcium carbonate carbonation experiment (row 1) and another experiment of a rock sample (row 2). In column A, no Poisson noise was generated, and in columns B-D, Poisson noise with blank scan factors of 100 was generated in the raw projection data to represent dose-limited experimental conditions. The actual COR, the CORs predicted by the proposed method and Vo's method are 1019.00, 1018.50, and 1065.50 pixels, respectively for the calcium carbonate experiment and 1024.00, 1023.50, and 1031.00, respectively for the rock experiment.}
\label{fig:stress_test_comparison}
\end{figure}

\begin{table}[ht]
\caption{Stress-test performance comparison between different COR estimation methods on the in-situ datasets.}
\smallskip
\begin{center}
\begin{tabular}{cccccc}
\midrule
Condition & Configuration & N & median (px) & IQR (px) & $\leq$ 2 px \\
\midrule
\multirow{2}{*}{Sparse: every $5^{th}$ projection}& Proposed & 10 & 0.50 & 0.88 & 10/10 \\
& Vo & 10 & 0.50 & 0.50 & 10/10 \\
\multirow{2}{*}{Sparse: every $10^{th}$ projection}& Proposed & 10 & 0.50 & 0.50 & 10/10 \\
& Vo & 10 & 0.50 & 0.88 & 10/10 \\
\multirow{2}{*}{Sparse: every $20^{th}$ projection}& Proposed & 10 & 2.00 & 1.25 & 7/10 \\
& Vo & 10 & 1.50 & 1.75 & 6/10 \\
\midrule
\multirow{2}{*}{Low-photon: blank scan factor 10}& Proposed & 10 & 0.25 & 0.50 & 10/10 \\
& Vo & 10 & 0.00 & 0.00 & 10/10 \\
\multirow{2}{*}{Low-photon: blank scan factor 100}& Proposed & 10 & 0.50 & 0.50 & 10/10 \\
& Vo & 10 & 8.25 & 39.25 & 5/10 \\
\multirow{2}{*}{Low-photon: blank scan factor 1000}& Proposed & 10 & 1.25 & 17.63 & 6/10 \\
& Vo & 10 & 12.75 & 38.00 & 3/10 \\
\midrule
\end{tabular}
\end{center}
\label{tab:stress_test_insitu}
\end{table}

\subsubsection{The second data source}
\paragraph{Accuracy} Table~\ref{tab:comp_tomobank} shows the absolute errors of the COR values estimated for the 8 datasets obtained from the TomoBank repository using the proposed method, the Vo's method and the SIFT method, respectively. The proposed method shows very similar accuracy to the Vo's method. In comparison, significantly larger errors are occasionally observed among the CORs estimated using the SIFT method due to key point detection failure. \par
\begin{table}[ht]
\caption{Absolute errors of the estimated CORs based on the proposed method, the Vo's method and the SIFT method on the TomoBank datasets.}
\smallskip
\begin{center}
\begin{tabular}{rrrrr}
\midrule
Configuration & $N$ & mean & median & max \\
\midrule
Proposed & 8 & 1.00 & 1.00 & 2.00 \\
Vo &  8 &   1.13 & 1.00 & 2.00 \\
SIFT &  8 &   3.80 & 0.97 & 22.39 \\
\midrule
\end{tabular}
\end{center}
\label{tab:comp_tomobank}
\end{table}
\paragraph{Stress tests} Table \ref{tab:stress_test_tomobank} reports the same stress test performed on the selected TomoBank datasets. Due to the higher diversity in the data and the wider search range of the algorithms, the errors generally show wider distributions than from the previous in-situ experimental dataset. As the number of projections decrease uniformly by factors of 2, the absolute errors of both methods remain low and largely unchanged until only every $10^{th}$ projection is available and increase substantially afterwards, similar to the trend observed from the previous in-situ data. In particular, the proposed method shows a consistently lower median absolute error and higher frequency of the absolute errors below 2 pixels than Vo's method. The proposed method also shows an IQR of the absolute errors no larger than those from the Vo's method for all sparse conditions except with only every $20^{th}$ projection available, in which case the proposed method shows a significantly larger IQR of the absolute errors. In terms of the blank scan factor, both methods show a sharp increase in the absolute errors when the underlying blank scan factor increases from 10 to 100. The proposed method also shows consistently lower absolute errors than Vo's method. \par
\begin{table}[ht]
\caption{Stress-test performance comparison between different COR estimation methods on the TomoBank datasets.}
\smallskip
\begin{center}
\begin{tabular}{cccccc}
\midrule
Condition & Configuration & N & median (px) & IQR (px) & $\leq$ 2 px \\
\midrule
\multirow{2}{*}{Sparse: every $5^{th}$ projection}& Proposed & 8 & 1.25 & 1.00 & 8/8 \\
& Vo & 8 & 2.00 & 1.00 & 7/8 \\
\multirow{2}{*}{Sparse: every $10^{th}$ projection}& Proposed & 8 & 1.00 & 0.75 & 7/8 \\
& Vo & 8 & 2.00 & 2.13 & 4/8 \\
\multirow{2}{*}{Sparse: every $20^{th}$ projection}& Proposed & 8 & 2.50 & 7.38 & 3/8 \\
& Vo & 8 & 3.25 & 2.88 & 2/8 \\
\midrule
\multirow{2}{*}{Low-photon: blank scan factor 10}& Proposed & 7 & 1.00 & 1.00 & 6/7 \\
& Vo & 7 & 1.50 & 2.50 & 5/7 \\
\multirow{2}{*}{Low-photon: blank scan factor 100}& Proposed & 7 & 1.50 & 32.75 & 5/7 \\
& Vo & 7 & 4.50 & 36.75 & 3/7 \\
\multirow{2}{*}{Low-photon: blank scan factor 1000}& Proposed & 7 & 9.50 & 26.75 & 2/7 \\
& Vo & 7 & 70.50 & 98.50 & 3/7 \\
\midrule
\end{tabular}
\end{center}
\label{tab:stress_test_tomobank}
\end{table}

\subsubsection{Takeaways}
\label{sec:comp_takeaways}

Three observations summarize the comparison:
\begin{enumerate}
  \item On the 302 out-of-distribution in-situ scans the proposed
        method agrees with the operator-validated Vo reference to
        within 2\,pixels on every scan, within 1\,pixel on 301
        (99.7\,\%) and within 0.5\,pixel on 277 (91.7\,\%). SIFT
        agrees with Vo at a similar sub-pixel level (within 2\,pixels
        on every scan). This validates the proposed method's
        robustness on sample chemistry not present in the training
        set; the dataset does not differentiate the three methods at
        the sub-pixel level.
  \item The same per-scan agreement holds across the full
        20--800\,$^{\circ}$C thermal program, on a specimen whose
        morphology and attenuation contrast evolve substantially
        under the carbonation reaction. The dataset is therefore a
        stringent test of robustness to sample-state variation at
        fixed instrument geometry, and the proposed method passes it.
  \item On both the in-situ dataset of calcium carbonate carbonation and the datasets from         TomoBank, the stress tests suggest the proposed method performs on par with the Vo's       method under the given reduction factors in the number of projections and                  significantly better than the Vo's method on noisy projection data.
\end{enumerate}

\section{Applications}
\label{sec:application}
\subsection{Deployment in tomocupy}
\label{sec:integration}

The classifier and its inference pipeline described above have been included in the latest release of the \emph{tomocupy} reconstruction package~\cite{nikitin2023tomocupy} and are already in production use at the APS of Argonne National Laboratory, where they are invoked daily on micro-tomography~\cite{Nikitin2023realtime} and nano-tomography~\cite{Deandrade2021fastnano} scans. In that integration the classifier is exposed through the \verb|--rotation-axis-method ai| option of the \verb|tomocupy recon| command, so that a COR sweep, the AI-based selection described in Section~\ref{sec:method}, and the final reconstruction can be issued from a single invocation; the slices passed to the classifier are obtained directly from \emph{tomocupy}'s in-memory reconstruction buffers, without the intermediate TIFF round-trip required by the
stand-alone package. The integrated path inherits the \emph{tomocupy} CUDA and HDF5 build-time dependencies, and is the preferred entry point when the data are already being reconstructed with \emph{tomocupy}. The stand-alone \emph{tomo-center} package described in this work remains the recommended form for reconstruction-back-end-agnostic deployments and for the per-beamline fine-tuning workflow. \par

\subsection{TomoGUI: a user interface for tomocupy}
\label{sec:tomogui}
\begin{figure}[h]
\begin{center}
\includegraphics[width=0.8\textwidth]{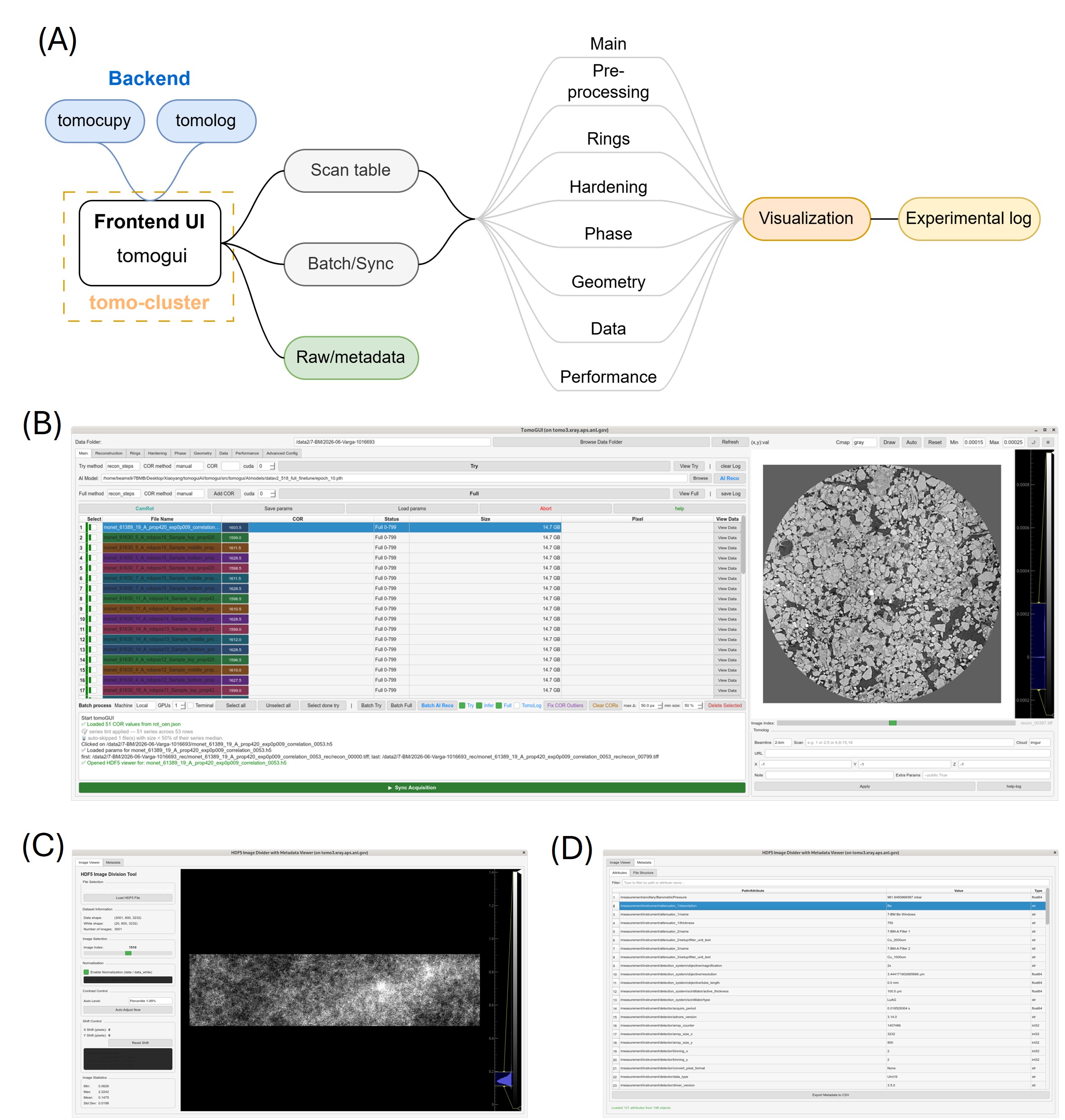}
\end{center}
\caption{The functions and user interface (UI) of tomogui. 
(A) The main functions and workflow. 
(B) The main UI. 
(C) The UI for visualizing RAW data. 
(D) The metadata tab of (C).}
\label{fig:tomogui_UI}
\end{figure}

To improve the user experience of tomocupy without command line interfaces, a graphical user interface (GUI) - TomoGUI is developed prioritizing the parameter selection, visualization, batch processing and automation. 
The GUI completely written with pyQT provides access to users all reconstruction parameters sorted out in different tabs. At the same time, it is equipped with tools for visualizing raw data stored in HDF5 format based on the Imaging Group data structure, as well as COR images and reconstructed datasets. As shown in Figure~\ref{fig:tomogui_UI} (A), TomoGUI provides a frontend interface connecting tomocupy for reconstruction and tomolog for experimental logging. The TomoGUI interface is organized into six functional regions to support parameter configuration, dataset management, batch or synchronized processing control, runtime monitoring, visualization, and tomolog setup (Figure~\ref{fig:tomogui_UI} B). Reconstruction parameters are organized into eight tabs with predefined default values. A scan table displays all datasets (*.h5) in the selected data directory. Files are automatically color-coded based on their prefix, improving visual differentiation and facilitating the identification of different datasets. The visualization panel enables viewing of reconstructed slices. In addition, the interface provides access to a separate data-viewing window that enables users to check raw projection images and associated metadata (Figure~\ref{fig:tomogui_UI} C-D).\par

TomoGUI supports single-scan, batch and synchronized processing modes. For the single-scan mode, after selecting the appropriate parameters, users can select the manual method to find the best COR, save the value in the scan table, and conduct a full volume reconstruction. Users can use the tomo-center AI method to find the COR and reconstruct the whole dataset automatically. The batch processing mode allows users to select datasets from the scan table and available machines in the tomo-cluster to run reconstructions. It is possible to configure the machine names and user accounts directly through the GUI, providing greater flexibility when the package is deployed on different systems. Lastly, the synchronized processing mode monitors the selected data directory and automatically initiates the reconstruction workflow when a new dataset is available. This mode integrates COR estimation using tomo-center, full-volume reconstruction, and tomolog generation, which significantly reduces the manual intervention. For each selected dataset folder, all center-of-rotation (COR) values and reconstruction parameters are automatically stored in a JSON configuration file located within the same directory. When a new reconstruction is launched with different parameters, the corresponding entries in the JSON file are automatically updated. If a configuration file is available, it is automatically loaded when a new TomoGUI session is started, ensuring continuity between sessions. TomoGUI is currently deployed and operational at all Imaging Group beamlines at APS.\par

\section{Discussion}
In this paper, we present a deep learning-based approach to classify tomographic slices reconstructed under the correct and incorrect COR parameters, and search the parameter space for the COR value that optimizes the tomographic reconstruction quality. Determination of the COR parameter for tomography experiments at the light source is an important step toward automating the full reconstruction pipeline and increasing the scientific value of synchrotron experiments for the user community. The disadvantage of prior automated methods is primarily the lack of robustness to variations in the imaging and sample conditions. In comparison, learning-based algorithms have been previously shown as a promising alternative to accurately detect the COR of tomographic measurements under actual acquisition conditions of synchrotron experiments. In this work, we further enhance the performance of the DNN by synergistically combining the highly generalizable feature extraction capability of DINO-pretrained ViT models and the highly efficient MIL framework to aggregate the most discriminative spatial features associated with the underlying COR reconstruction parameter and determine its optimal value. This proposed method is accurate, robust to typical imaging noise during the operation, generalizable to different samples, and adaptive to changing sample conditions. Moreover, it has been integrated into several tomography software packages to assist experiments during the routine beamline operations.\par

Given a series of tomograms from a typical try reconstruction, the key technical challenge to searching the optimal COR parameter is the algorithmic efficiency. Since synchrotron tomography typically has a very high spatial resolution, the underlying reconstruction pipelines can involve a computational workload operating at over gigapixel-scale throughput in the joint spatial–parameter space. In this context, learning-based algorithms need to be designed to reduce the input information to keep the resulting inference workload manageable by the runtime system during the routine user operation. For one thing, the ViT backbone pretrained under the DINOv2 framework has learned to encode very complex intensity distributions through exposure to large-scale, well-curated natural image datasets. Adapting such a backbone model to tomographic data further ensures effective transfer of general visual knowledge to extract and represent specific tomographic image features in the absence of large-scale in-domain training data. For another, the adoption of a classic end-to-end MIL framework for both training and inference considerably reduces the computational complexity associated with the use of a ViT, allowing the use of image patches in place of the entire tomograms without substantially increasing the risk of overfitting. As a result, the proposed method effectively balances algorithm accuracy and efficiency, and can be deployed in production environment to support the operations.\par

In order to interpret the importance of different patches of the input tomogram to the classification of its underlying COR by the proposed model once it was trained, the attention scores were extracted from the gated attention stage of the inference pipeline and visualized as a heatmap, similarly as in previous studies \cite{ilse2018attention,shao2021transmil}. According to the results, the foreground region of the reconstructed tomogram has a significantly higher score than the background, consistently with different experiments and across distinct COR reconstruction parameters. This is expected since the incorrect COR-induced double-edge and tangential-streak artefacts are primarily concentrated within the foreground region and in its vicinity. In other words, once the model has learned to attend to the foreground region of the tomogram, the resulting features due to the misspecification of COR become more straightforward to extract. It should be noted that such high interpretability of the proposed method may also be the result of the synergy between the DINO pretraining framework and the ViT model architecture, which has been demonstrated to explicitly capture the semantic segmentation information of an image \cite{caron2021emerging}. In this work, to ensure success of the classification task, a minimum window size was maintained which limited the resolution of the attention map. More fine-grained attribution of spatial features to the classification is left for future investigation. \par

We already use the proposed methods on a regular basis through their integration within TomoGUI. One example is nano-tomography datasets acquired using X-ray Absorption Near Edge Structure imaging at the APS 32-ID beamline. Coupled with synchronized reconstruction workflows, this methodology enabled the automatic processing of more than 2,000 scans with minimal human intervention. Another example includes micro-tomography acquisitions carried out at the 7-BM beamline on soil core samples with different sizes and collected from different fields. The method successfully reconstructed for more than 200 scans as well. \par 

We now discuss areas where the proposed method may be further improved in the continuous development roadmap. First, the ViT backbone architecture has a known computational complexity that scales quadratically with the image size in 2-D. As a result, enlarging the patch size may significantly increase the inference time. In comparison, multiple input patches are independently batch-processed by the backbone model, and the influence of the number of patches on the inference time is hence less pronounced. In the selected configuration, we used 3 patches of the size $518\times 518$ and achieved sub-pixel accuracy in the COR estimation based on the testing datasets. We noticed that the use of smaller or fewer patches may reduce the accuracy. For example, with a single patch of size $224\times 224$, there could be an increasing probability of missing discriminative features of the COR-induced artefacts (i.e., the patch may cover the background region), leading to false positives. Further reducing the patch size may risk losing distinct features associated with the specific image classes. In order to reduce the inference time without compromising the accuracy, the proposed method may be applied as a refinement step to classify fewer candidate COR parameters following an initial coarse search, such as via the Vo's or other methods. Alternatively, given the consistency of the attention map to various COR parameters, the attention map may be used to constrain the configuration of the patch size and number and improve the search efficiency. These options may be explored in the future under the framework of an agentic system to intelligently navigate a variety of tomographic experiments.\par

Second, the training dataset compiled for this work represents a small dataset from retrospective studies conducted at several tomography beamlines at the APS. With this dataset, our intention was to quickly establish a data-driven algorithmic framework and deliver a practical tool to support the beamlines' routine operations. At the training time, these data were not augmented with the sparsity or dose constraints typical of those used in the specific stress tests, and other challenging situations during the synchrotron tomography experiments, such as missing wedges and double field-of-views, were also not explicitly considered. In our past work on high-speed imaging, we noticed that task-specific data augmentation effectively improved model performance in related downstream applications \cite{tang2025deep,tang2025swin}. In addition, the limited testing data were also not meant as a comprehensive test to establish the proposed method as a replacement of other prior methods, despite the great potentials of the proposed method that can be noted from these tests. Nonetheless, the trained model, once integrated into the beamline software, was found to be capable of handling most experimental conditions seen so far with accurate COR predictions. As such, we will continue to grow the training dataset and improve the model generalizability as part of the regular maintenance of the beamline infrastructure to keep pace with evolving user needs during ongoing experiments. Techniques such as one-shot learning and test-time adaptation may also be explored to extend the trained model to images with distinct distributions. These potentials will be investigated in the future.\par

\section{Conclusions}
In this paper, we present \emph{tomo-center}: a deep learning-based algorithm to calibrate the rotation axis of a synchrotron tomography instrument by analyzing features of the reconstructed tomograms with correct and incorrect COR parameters. The algorithm is implemented in Python and publicly available at \url{https://github.com/xray-imaging/tomo-center/releases/tag/v1.0.0}. MAEs of the proposed method were tested to be consistently within one pixel based on two independent data sources, suggesting high estimation accuracy and good model generalization. The proposed method was also tested to operate robustly when the available number of tomographic projections was reduced by a factor of up to 10 or the blank scan factor of the underlying Poisson's noise was increased to 10. In addition, we demonstrated the trained DNN model was able to attribute the correct classification to the right spatial region, in alignment with human visual perception.\par
We also present the integration of the proposed methodology into \emph{TomoGUI}, a graphical user interface designed to streamline tomography data processing and reconstruction workflows at synchrotron facilities. TomoGUI provides tools for data management, visualization, COR determination, and reconstruction, facilitating routine deployment of advanced processing methods in a user-friendly environment. The software is currently operational across all Imaging Group beamlines at the Advanced Photon Source (APS) and publicy available at \url{https://github.com/xray-imaging/tomogui}.\par

In summary, the proposed algorithm and its integration into TomoGUI provide a practical and robust solution for automated COR calibration, supporting high-throughput tomography experiments with minimal user intervention.

\begin{acknowledgements}
The authors thank the DINOv2 team at Meta AI Research for the pretrained
vision-transformer backbone.
\end{acknowledgements}

\begin{funding}
This research used resources of the Advanced Photon Source, a U.S.
Department of Energy (DOE) Office of Science User Facility operated for the
DOE Office of Science by Argonne National Laboratory under Contract
No.~DE-AC02-06CH11357. This research is based
on work supported by Laboratory Directed Research and
Development (LDRD) funding from Argonne National Laboratory,
provided by the Director, Office of Science, of the U.S. DOE
under Contract No. DE-AC02-06CH11357. This research used
resources of the Argonne Leadership Computing Facility, a U.S.
Department of Energy (DOE) Office of Science user facility
at Argonne National Laboratory and is based on research
supported by the U.S. DOE Office of Science, Advanced Scientific
Computing Research Program, under Contract No.~DE-AC02-06CH11357. Part of the research is supported by eBERlight program, funded by the U.S. DOE office of Biological and Environmental Research (BER).
\end{funding}

\ConflictsOfInterest{The authors declare no conflicts of interest.}

\DataAvailability{The \emph{tomo-center} source code is available at
\url{https://github.com/xray-imaging/tomo-center/releases/tag/v1.0.0} under a BSD-3 license. The
classifier checkpoint is distributed by Argonne National Laboratory at
\url{https://anl.box.com/s/k85a89kyplzd56hnjudw4ergt6ojfhoa}.

The \emph{TomoGUI} source code is available at
\url{https://github.com/xray-imaging/tomogui} under a BSD-3 license.

The example datasets used in Section~4 are available from the corresponding author
upon reasonable request.}
\bibliography{refs}

\end{document}